\documentclass[reprint,superscriptaddress,showpacs,amsmath,amssymb,aps,pre]{revtex4-1}
\usepackage{multirow}
\usepackage[caption=false]{subfig}
\usepackage{float}
\usepackage{amssymb,amsbsy,color}
\usepackage{graphicx}
\usepackage{dcolumn}
\usepackage{bm}
\usepackage{txfonts}

\begin{document}

\title{Building blocks of the basin stability of power grids}
\author{Heetae Kim}
\affiliation{Department of Energy Science, Sungkyunkwan University, Suwon, 16419, Korea.}
\author{Sang Hoon Lee}
\affiliation{School of Physics, Korea Institute for Advanced Study, Seoul, 02455, Korea.}
\author{Petter Holme}
\email{holme@skku.edu}
\affiliation{Department of Energy Science, Sungkyunkwan University, Suwon, 16419, Korea.}

\date{\today}

\begin{abstract}
Given a power grid and a transmission (coupling) strength, basin stability is a measure of synchronization stability for individual nodes. Earlier studies have focused on the basin stability's dependence of the position of the nodes in the network for single values of transmission strength. Basin stability grows from zero to one as transmission strength increases, but often in a complex, nonmonotonous way. In this study, we investigate the entire functional form of the basin stability's dependence on transmission strength. To be able to perform a systematic analysis, we restrict ourselves to small networks. We scan all isomorphically distinct networks with an equal number of power producers and consumers of six nodes or less. We find that the shapes of the basin stability fall into a few, rather well-defined classes, that could be characterized by the number of edges and the betweenness of the nodes, whereas other network positional quantities matter less.
\end{abstract}
\maketitle

\section{\label{sec:intro}Introduction}
A stable supply of electricity is crucial for our society. Finding network topologies that enhance the stability of power grids has been an active subfield of network theory. Previous studies have typically focused on the structural vulnerability of power grids against external attacks~\cite{PhysRevE.69.025103,RISA:RISA791,Chaos.19.1,Bompard20095,RosasCasals:2013jo,Dobson:2007fq,Chen2010595} and the cascading spreading of system failure over power grids~\cite{PhysRevE.69.045104,EPJB.46.1,Dobson:2007fq,Pahwa:2014ho}.
In addition, researchers have studied the relation between topology and dynamics via network synchronization models. Filatrella \emph{et al.}~\cite{Filatrella:2008co} derived a synchronization model for power-grid nodes as a form of the Kuramoto model. Other papers investigated the relationship between the network topology and synchronization of power grids~\cite{Motter:2013iw,Dorfler:2013ew,Rohden:2014fy,Nishikawa:2015gl}. 

Recently, Menck \emph{et al.}~\cite{Menck2013basin} introduced the concept of \emph{basin stability} (denoted by $B$ in this paper) quantifying a type of stability of synchronization in power-grid nodes. In a general nonlinear system, the basin stability of a given
stable state measures the likelihood that the system recovers the state against an external perturbation. For power grids, the basin stability of a node can be defined as the fraction of initial phase angle and frequency values of that node which result in convergence to synchronization according to the system dynamics (with the initial values fixed for all other system variables). Since calculating basin stability requires the Monte Carlo sampling of nonlinear systems, it is computationally expensive. Accordingly, it is suggested to use the connection-topology-based identification method---namely, the topological characteristics such as dead-end, dead-tree, and detour---to identify nodes with large or small basin stability values for computational tractability~\cite{Ji:2014in,Menck:2014fn,Schultz2014detours}. They exploit the correlation between basin stability and topological characteristics of nodes. 

A topic that has gotten relatively less attention is the effects of transmission strength $K$ on the basin stability. Transmission strength greatly affects nodes' basin stability, as shown in our recent study that reveals that power-grid nodes exhibit various patterns of basin stability transition as a function of the transmission strength~\cite{BasinStability_NJP}. Since the transition pattern is a consequence of the nontrivial combination of connection topology and parameters for power-transmission dynamics, different transition patterns of nodes can effectively represent each node's characteristics in terms of dynamical behaviors.

In this study, we systemically analyze the transition of basin stability. We generate all of the possible configurations of small networks of given sizes, and investigate the transition patterns. We find specific subregions to which the basin stability of nodes are concentrated. Based on the trajectory of the basin stability transition, we classify nodes into different clusters, which cannot be explained by conventional clustering methods.

\section{\label{sec:background}Basin stability transition}
\subsection{\label{diffent pattern}Various transition patterns}
Quantities characterizing power-grid systems such as the amount of power generation or consumption, phase difference of nodes, connection structure of the power grid, and coupling strength affect the basin stability of nodes. The governing dynamics for the synchronous interaction between power-grid nodes is often described as the second-order Kuramoto-type model~\cite{RevModPhys.77.137,Ji:2014in,Motter:2013iw,Menck2013basin,Menck:2014fn,Schultz2014detours,Nishikawa:2015gl}:
\begin{equation}
\ddot \theta_i = \dot \omega_i = P_i - \alpha_i \dot \theta_i - \sum_{i \ne j} K_{ij} \sin(\theta_i - \theta_j) ,
\label{eq:Kuramoto_type_equation}
\end{equation}
where $P_i$ is the net power generation or consumption at power grid node $i$, $\alpha_i$ is the dissipation constant, $K_{ij}$ is the coupling strength between nodes $i$ and $j$, $\theta_i$ is the phase angle of node $i$, and $\omega_i = \dot \theta_i$ is the angular velocity of node $i$. The phase angle and frequency are measured relative to the reference frame. When $\omega_i$ vanishes for all of the nodes, the power grid is considered to be synchronized and the system maintains the constant frequency desired for stable power transmission. 

\begin{figure}
\begin{center}
\includegraphics[width=0.9\linewidth]{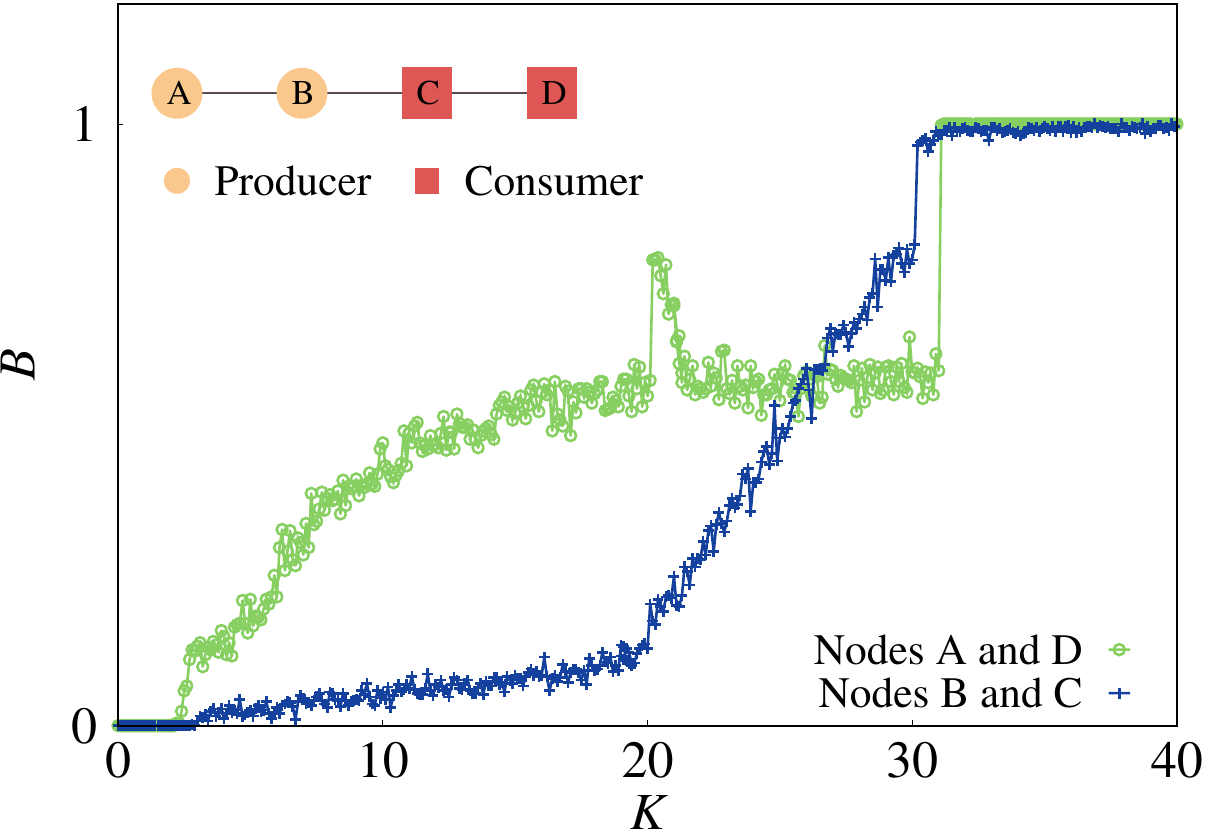}
\end{center}
\caption{(Color online) Basin stability transition of nodes as a function of the coupling strength $K$ in a network. Node $i$ ($\in \{A, B, C, D\}$) is net producer (consumer) when $P_i > 0$ ($P_i < 0$), respectively. In spite of the same functional role of nodes $A$ and $B$ in the linear chain network, the stability transition patterns of them are very different. Node $C$ ($D$) shows the same pattern with node $B$ ($A$), respectively.}
\label{fig_1}
\end{figure}

The transition pattern of basin stability varies across different nodes. It is known that the basin stability nonmonotonically changes as the coupling strength increases~\cite{BasinStability_NJP}. Accordingly, the functional form of the transition pattern can also be diverse for nodes. In a network depicted in Fig.~\ref{fig_1}, four nodes are connected in a row. The parameter values such as the amount of input or output $|P|$ and the dissipation constant $\alpha$ of the four nodes for the synchronization dynamics are set to be the same with each other, except for the fact that the two nodes on the left are producers and the other two nodes on the right are consumers(see Sec.~\ref{sec:simulation} for the detailed parameter setting.). Despite the fact that nodes $A$ and $B$ ($C$ and $D$) play the same role as producers (consumers), respectively, their stability transition patterns are different. Node $A$ seems to maintain an intermediate level of basin stability before it suddenly reaches the maximum (unity) for large coupling strength. On the other hand, node $B$ maintains a low level of basin stability up to $K \simeq 20$ and then it rapidly increases. It is interesting to note that producers and consumers show the same transition pattern when their network-topological location is the same. In other words, one can get the same transition pattern from the parameter setting for the opposite sign of $P$, i.e., in Fig.~\ref{fig_1}, node $C$ ($D$) has the same transition pattern with node $A$ ($B$), respectively.

\subsection{\label{sec:simulation}Numerical simulation}
For numerical integration of Eq.~\eqref{eq:Kuramoto_type_equation}, we use the same method used in Ref.~\cite{BasinStability_NJP}. We assume the same strength value $K$ for all of the transmission lines, so that $K_{ij}=K$ if there is an edge between nodes $i$ and $j$ and $K_{ij}=0$ otherwise. The amount of input power of node $i$, or $P_i = 1$ if node $i$ is the net power producer and $P_i = -1$ if it is the net power consumer. Following Refs.~\cite{Menck2013basin,Menck:2014fn,Schultz2014detours}, we set the dissipation constant $\alpha=0.1$ for all of the nodes. The initial random perturbation is assigned for given node $i$ as $-100\leq\omega_i\leq100 $ and $-\pi\leq\theta_i<\pi$. Besides the node $i$, the other nodes $j \neq i$ have the initial phase $\theta_j=0$ and angular velocity $\omega_j=0$. We consider the system is synchronized and stable when the system converges with the numerical convergence criteria~\cite{Menck:2014fn,BasinStability_NJP}, namely, the time derivative of angular frequency $\dot\omega < 5\times10^{-2}$ and the angular frequency $\omega < 5\times10^{-2}$ for all of the nodes.

\subsection{\label{sec:flow}Flow betweenness}
We introduce \textit{flow betweenness} to calculate how much a node functionally contributes particularly for the power-grid interactions. Newman~\cite{Newman2005current} suggested the \emph{current flow betweenness centrality} as a centrality measure  inspired by the current flow of an electric circuit. Schultz \textit{et al}.~\cite{Schultz2014detours} successfully applied the centrality measure to identify basin stability of nodes in a power grid. 
The original current flow betweenness considers all of the node pairs regardless of the node identities: producers ($P > 0$) or consumers ($P < 0$). 

In this study, however, to take the distinction between producers and consumers into account, we modify the original method to explicitly deal with the relevant source-target pairs.
In a real power grid, multiple numbers of producers and consumers exist through which current is injected into the power grid or extracted out of it.
Therefore, taking that distinction into account could result in the more realistic load distribution originated by the actual current flow, than the all-to-all pairs. 
In addition, the modified method requires a shorter computation time. Therefore, flow betweenness is particularly useful for large size power grids, where the computational speed becomes an issue.

The current flowing through an edge between two nodes is proportional to the voltage difference between them, which is affected by their locations. We calculate the flow betweenness of nodes in a network which has multiple sources and targets.
Let $\mathbf{u}$ denote the source vector whose $i$th element is $+1$ if $i$ is a source, and $-1$ if it is a target, and $0$ otherwise. The roles of nodes are identified by aforementioned $P_i$. 

The voltage vector $\mathbf{v}$ can be calculated from the admittance matrix $\mathbf{Y}$ and source vector $\mathbf{u}$ because $\mathbf{Y}\mathbf{v}=\mathbf{u}$ (matrix product). In the network representation of a power grid, the Laplacian matrix $\mathbf{L}$ of the network corresponds to the admittance matrix $\mathbf{Y}$ of the power grid. Therefore, the voltage vector $\mathbf{v} =\mathbf{Y}^{-1}\mathbf{u}=\mathbf{L}^{-1}\mathbf{u}=\mathbf{T}\mathbf{u}$, where $\mathbf{T}$ is the Moore-Penrose pseudo-inverse~\cite{Newman2005current, Schultz2014detours} of $\mathbf{L}$. Accordingly, the voltage at node $i$ with given source nodes $s$ and target nodes $t$ is 
\begin{equation}
v_i=\sum_{s}T_{is}-\sum_{t}T_{it}.
\label{eq:voltage}
\end{equation}

The current flowing through the edge between nodes $i$ and $j$ is proportional to the sum of the voltage difference between $i$ and $j$:
\begin{equation}
I_{ij} = C_{ij}\left[v_i-v_j\right],
\label{eq:current}
\end{equation}
where $C_{ij}$ corresponds to the conductance of the transmission line between nodes $i$ and $j$. Here we set $C_{ij}=A_{ij}$ (the adjacency matrix elements) for all of the edges. Note that we construct the electric power grid as an unweighted and undirected network, assuming that the conductance of all of the transmission lines is equal. $I_{ij}>0$ represents that the current flows from $i$ to $j$ and $I_{ij}<0$ means the opposite direction.

The current flow load at node $i$ is given by the sum of the current load from the connected edges to node $i$. We consider both throughput and input currents for calculation. Finally, the flow betweenness of node $i$ is
\begin{equation}
F_i = \frac{1}{2}\bigg(\left|u_i\right|+\sum_{j} A_{ij}\left|I_{ij}\right|\bigg) \,,
\label{eq:tcfb}
\end{equation}
where the source vector component $u_i$ is input and $\sum_{j} A_{ij}\left|I_{ij}\right|$ corresponds to the throughput of $i$. We normalize $F_i$ by 2 to avoid double counting.
We use unit-based real numbers for parameter values for the sake of simplicity as DC circuit approximation.

\section{\label{sec:result}Result}
\subsection{\label{motif}Two-node and four-node networks }
The shape of the basin stability transition is the synchronization stability landscape that has information about the collective dynamics of power-grid synchronization. Therefore, it is beneficial to design or operate the power grid with the knowledge of the factors that decide the transition patterns.
In our previous work~\cite{BasinStability_NJP}, we reveal that the mesoscale network property given by community consistency is an important factor.
For a more systematic approach from even simpler structures, we start from small networks by enumerating the exhaustive list of possible network configurations. We generate all possible network configurations with two, four, and six nodes (only even numbers to make the network balanced---the equal numbers of producers and consumers and their amount of production and consumption is the same). Since all of the nodes in real power grids should be connected, we only consider connected network configurations. 

\begin{figure*}
\begin{center}
\includegraphics[width=0.8\textwidth]{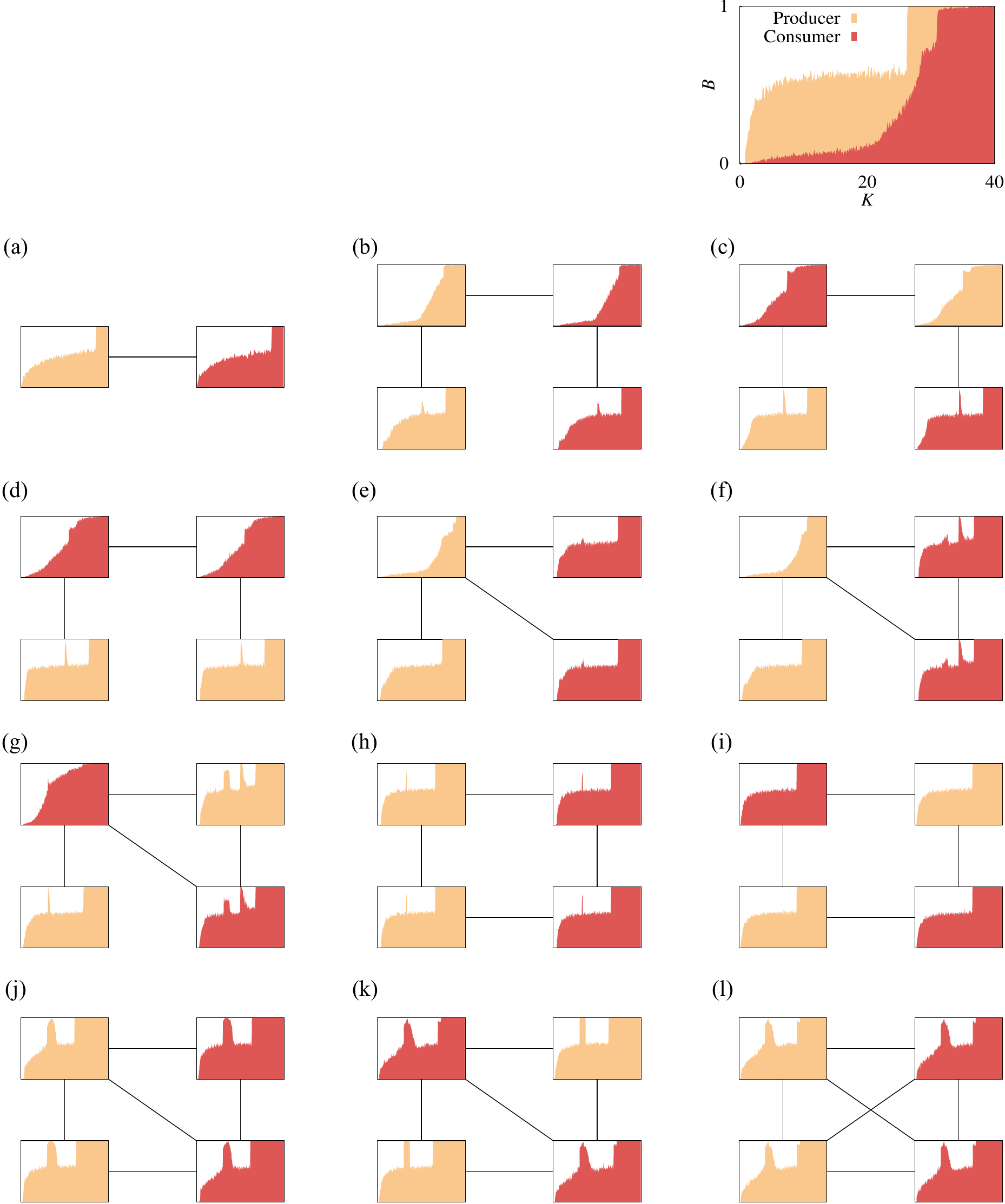}
\end{center}
\caption{(Color online) Basin stability transition patterns of nodes from all of the unique network configurations for two and four nodes. Each panel represents a node and its basin stability transition. Orange (bright) nodes play producers and red (dark) nodes are consumers. The edges between nodes are transmission lines connecting two nodes. The upper right panel above panel (c) is an illustration for legends. (a) is the network work motif with two nodes and (b)--(l) are that of four nodes. The number of effectively isomorphic configurations is (a) 1, (b) 23, (c) 23, (d) 23, (e) 23, (f) 23, (g) 47, (h) 11, (i) 5, (j) 23, (k) 11, and (l) 5. (a) shows $0\leq K \leq150$ and (b)---(l) have $0\leq K \leq40$ for $K$ range at X axis.}
\label{fig_2}
\end{figure*}

The number of network combinations with two different attributes enormously increases as the number of nodes increases. In order to effectively select only unique configurations up to network isomorphism and producer-consumer symmetry, isomorphic networks, which are identical to other networks after switching node attributes or node identity, are excluded from the configuration lists. As a result, among $2$ and $228$ cases for two- and four-node combinations, we identify $1$ and $11$ unique network configurations (Fig.~\ref{fig_2}), respectively.

In the simplest two-node network, the transition patterns of the two nodes are similar to each other. The basin stability gradually increases as $K$ increases and reaches the maximum (unity) at large $K$ [Fig.~\ref{fig_2}(a)], which is the same as the infinite-bus-bar model (for a node of interest, taking all of the other nodes as the environment) transition of basin stability as a function of coupling strength~\cite{Menck:2014fn}. 

\begin{figure}
\begin{center}
\includegraphics[width=0.9\linewidth]{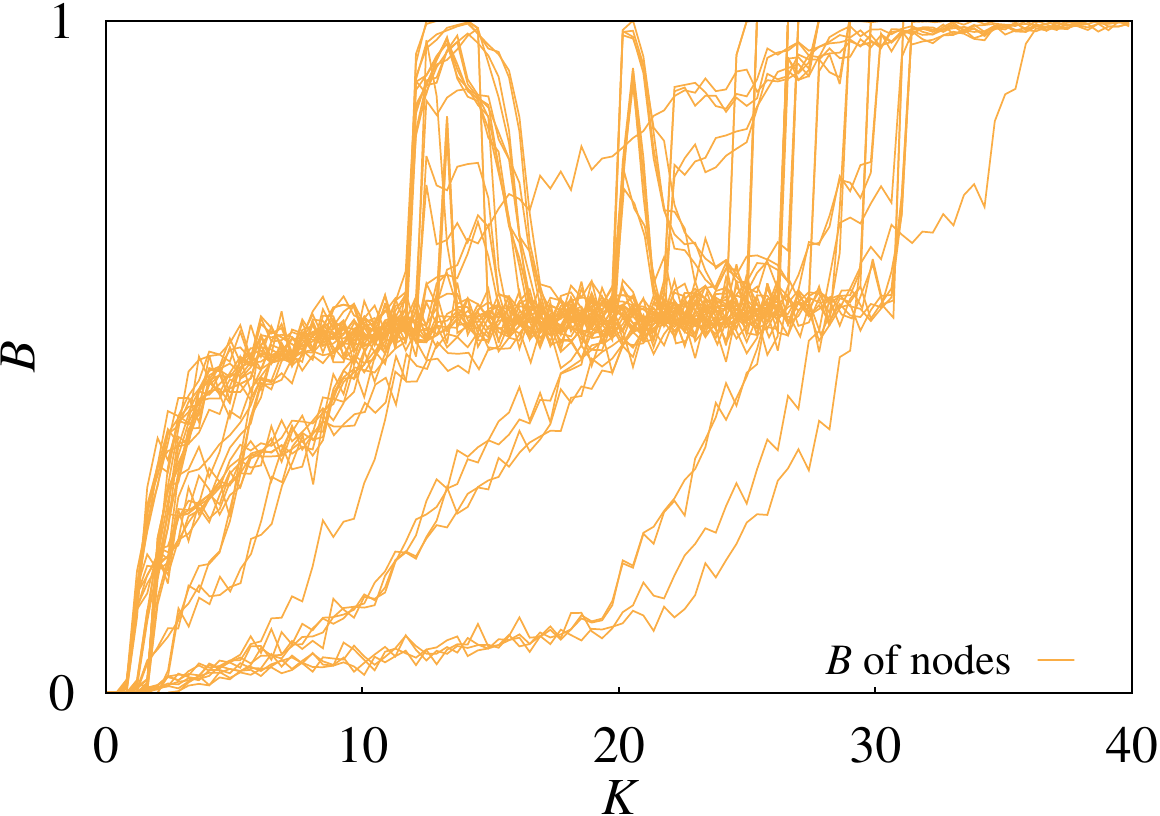}
\end{center}
\caption{(Color online) Basin stability transition of some nodes from two-node and four-node networks in Fig.~\ref{fig_2}. Various shapes of transition patterns exist in the network motifs.}
\label{fig_3}
\end{figure}

However, four-node networks show heterogeneous patterns for transition curves [Figs.~\ref{fig_2}(b)--~\ref{fig_2}(l)].
Some features of the transition shape from the four-node network are found. At first, various shapes of transition patterns appear. We classify them as the suppressed form [e.g., the top left node of Fig.~\ref{fig_2}(b)], enhanced form [e.g., the top left node of Fig.~\ref{fig_2}(g)], infinite-bus form [e.g., the bottom left node of Fig.~\ref{fig_2}(i)], and broad peak [e.g., the bottom left node of Fig.~\ref{fig_2}(j)].

The suppressed form and the enhanced form are identified comparing to the infinite-bus form. The infinite-bus form is characterized by the prolonged intermediate level of the basin stability region before the basin stability suddenly reaches unity. It is observed in the network with two nodes where one is an oscillator and another is an infinite bus. The suppressed form is identified by the lower rate of basin-stability increase as a function of the $K$ value than the one of infinite-bus form mostly at small $K < 20$. On the other hand, the enhanced form shows the higher rate of basin stability increase than the infinite-bus form without the intermediate stable basin stability region. The broad-peak form characterizes the unique feature of some nodes that the node shows the temporal maximum basin stability before it completely shows unity at large $K$.

Second, both the network structure and the position of producers and consumers collectively affect the basin stability transition. 
The different network topology could accompany the different transition pattern of basin stability, since the connection pattern could be different.
Yet, the different network topology could also generate the similar transition pattern of basin stability such as the networks in Figs.~\ref{fig_2}(k) and ~\ref{fig_2}(l).
The influence of the nodal attribute position on the basin stability transition also varies. For instance, the networks with the same topology seem to show similar transition patterns regardless of the position of producers and consumers as seen in Figs.~\ref{fig_2}(c) and ~\ref{fig_2}(d), ~\ref{fig_2}(h) and ~\ref{fig_2}(i), and ~\ref{fig_2}(j) and ~\ref{fig_2}(k). 
However, the different position of producers and consumers also results in the different transition shape such as the networks in Figs.~\ref{fig_2}(f) and ~\ref{fig_2}(g), particularly for the upper left nodes.
Considering the various consequences of the position of producers and consumers and the network structure for the transition pattern of basin stability, the collective influence of the nodal position and the network structure on the transition pattern needs further investigation, which we will extensively address in Sec.~\ref{sec:six_nodes}.

Third, when a node belongs to a triangle loop, the transition peak---temporary increase of basin stability---appears before the synchronous dynamics reaches unity [Figs.~\ref{fig_2}(f), ~\ref{fig_2}(g), ~\ref{fig_2}(j), ~\ref{fig_2}(k), and ~\ref{fig_2}(l)].
When the triangle participant has another neighbor with only a single neighboring node, the transition shape of the triangle participant becomes extremely biased to be suppressed---the top left node of Fig.~\ref{fig_2}(f)---or enhanced---the top left node of Fig.~\ref{fig_2}(g).

Last, when the number of edges is larger than four, all of the nodes show the broad transition peak shape on their transition form [Figs.~\ref{fig_2}(j)--~\ref{fig_2}(l)]. In other words, the peak shape appears when the network is dense.

\subsection{\label{sec:six_nodes}Six-node network motifs}
\subsubsection{\label{sec:3d}Clustering in the three-dimensional parameter space}

\begin{figure}
\begin{center}
\includegraphics[width=0.9\linewidth]{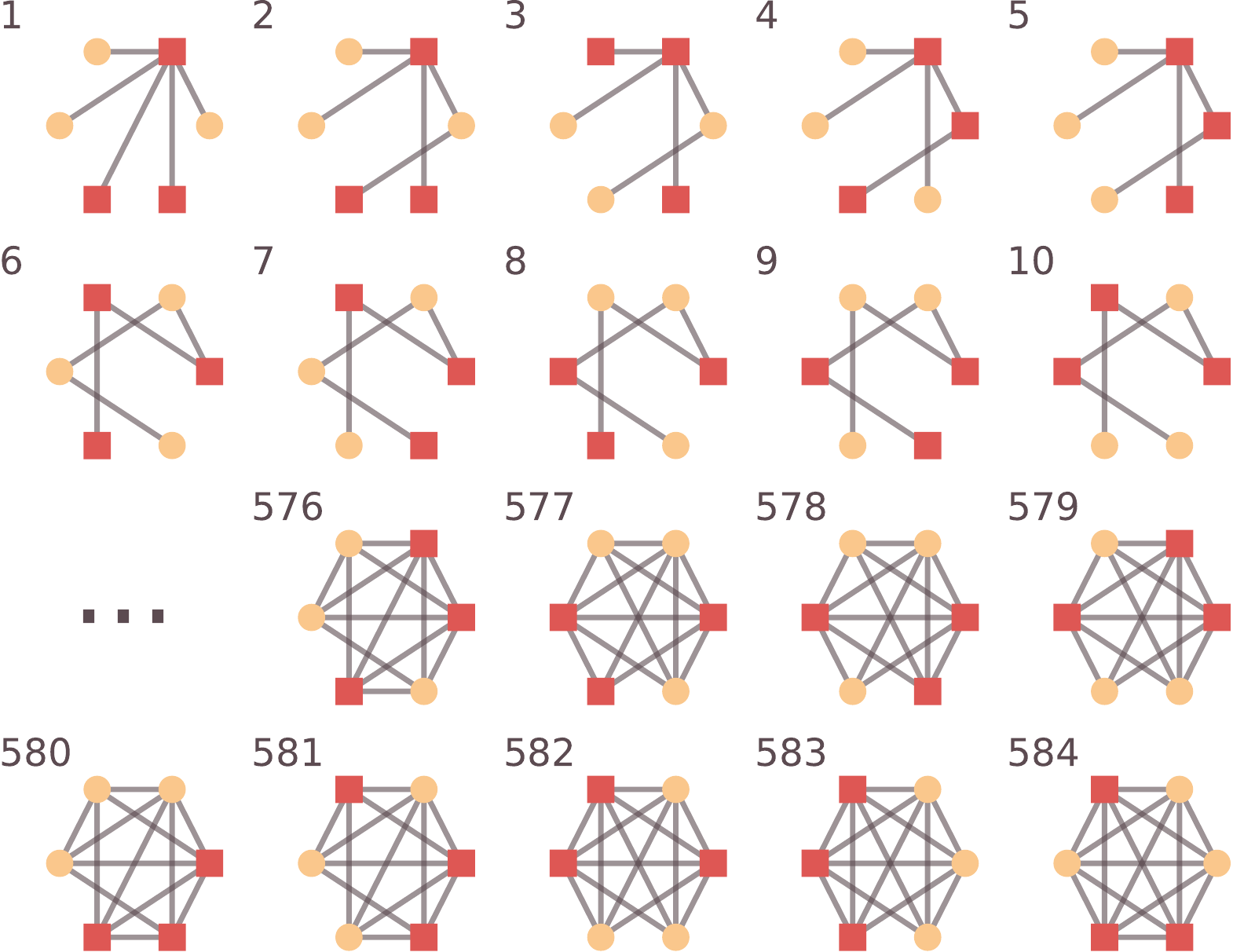}
\end{center}
\caption{(Color online) The network motifs with six nodes. Orange circle nodes (bright) are producers and purple square nodes (dark) are consumers. With six nodes, the number of unique network motifs is 584. The exhaustive list of configurations is shown in Ref.~\cite{SM}.}
\label{fig_4}
\end{figure}

We now consider the networks with six nodes. With six nodes, there are $584$ unique configurations when we take the isomorphism into account, from the $30827$ combinations (Fig.~\ref{fig_4}; see Supplemental Material~\cite{SM} for the detailed topology for all of the configurations.). As we observed in the four-node configurations, the shape of the basin stability transition varies for nodes. Since measuring the basin stability requires numerical simulations which are too computationally intensive, we choose to analyze the three-dimensional (3D) projection of the basin stability at three different $K$ values in order to effectively investigate the transition patterns without the exhaustive enumeration. Since the basin stability of nodes increases from zero at small $K$ to unity at large $K$, the transition pattern is effectively distinguished by comparing a couple of basin stability values at different $K$. We choose three different $K$ in our study. Then, the position of a plot in the 3D parameter space (where each axis corresponds to each $K$ value) provides the information about to which pattern the transition belongs. We use $K=7$, $14$, and $21$ values to represent the transition pattern of a node.

\begin{figure}[hbp!]
\begin{center}
\includegraphics[width=0.9\linewidth]{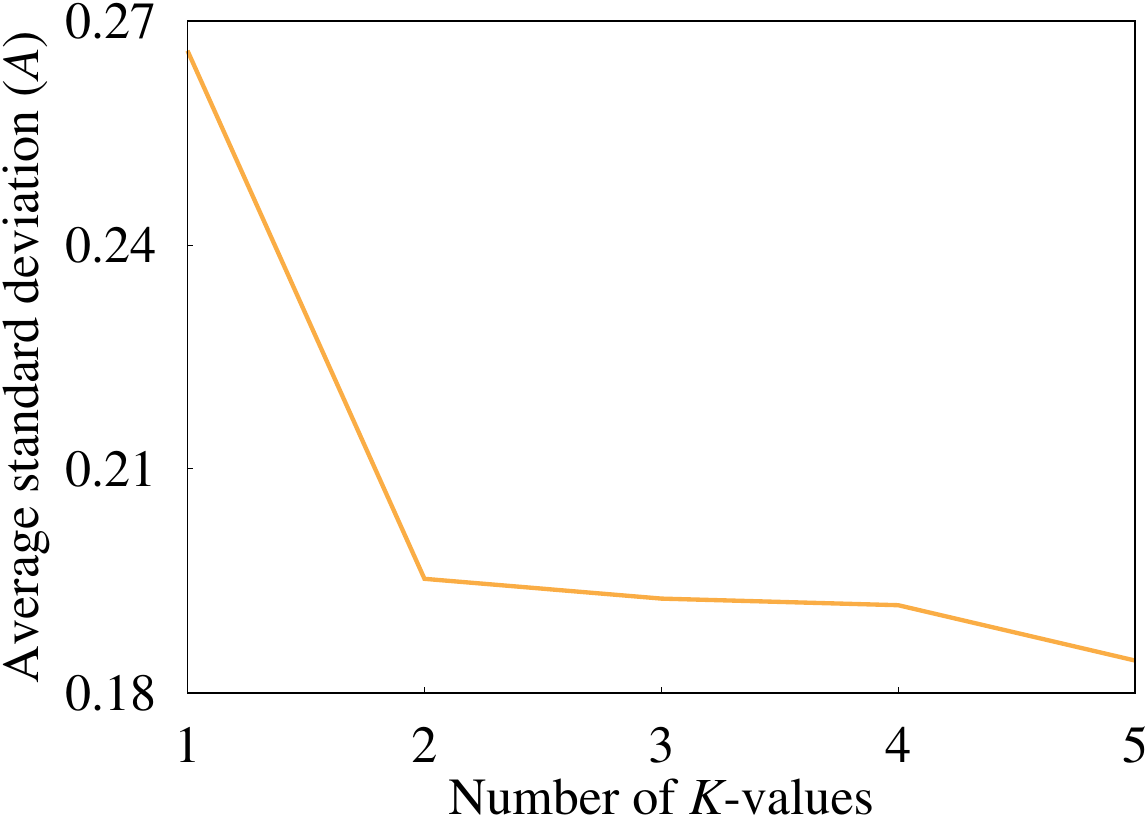}
\end{center}
\caption{(Color online) The average standard deviation of basin stability of nodes in each clustering for different numbers of $K$ values. It shows that the significant advantage of using three $K$ values for classification rather than only one $K$ value.}
\label{fig_5}
\end{figure}

Using three $K$ values can describe the shape of the transition curves much more accurately than using only a $K$ value. Here we show the deviation of basin stability in each group (a clustering of nodes) becomes sufficiently small with three $K$ values. 
To compare the different dimensional classifications, we classify curves into ten groups based on the basin stability of nodes at a given number of $K$ values (1 to 5). We then evaluate the goodness-of-classification of the clustering at all 28 $K$ values for all nodes in all six-node networks. The goodness-of-classification metrics is the root mean square of the deviation from the group mean, i.e.,
\begin{equation}
A=\sqrt{\frac{1}{n_Km}\sum_{K}\sum_{i}[B_i(K)-B_{g(i)}(K)]^2},
\end{equation}
where $n_K$ is the number of $K$ values considered, $m$ is the number of curves (i.e., the number of nodes in unique positions in all the six-node networks), $g(i)$ is the group $i$ belongs to, and $B_g(K)$ is the average value of $B$ at coupling strength $K$ for nodes in group $g$.
The result in Fig.~\ref{fig_5} implies that increasing only a few more $K$ values can make the classification much more accurate and completes our argument.

\begin{figure*}
\includegraphics[width=0.9\linewidth]{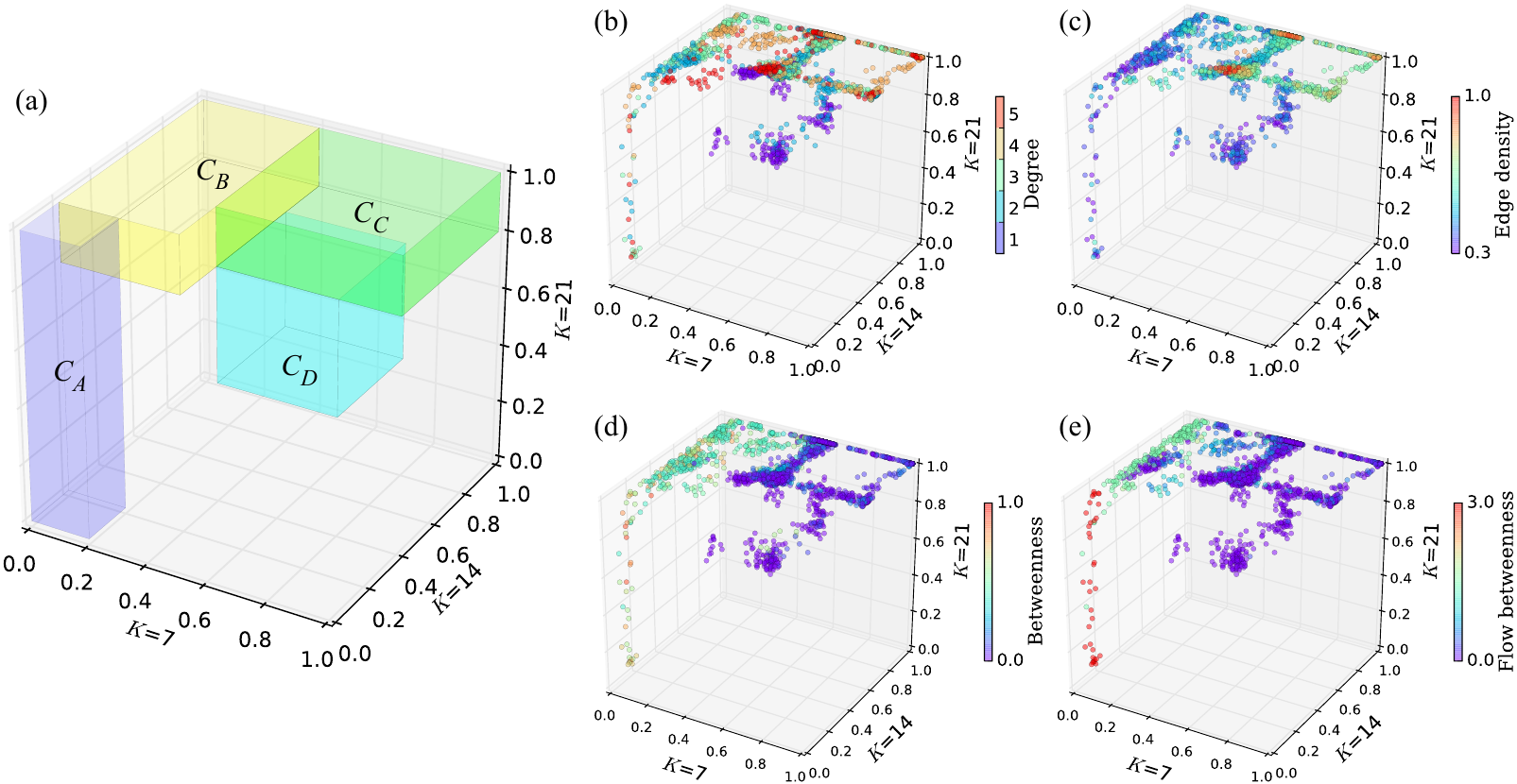}
\caption{(Color online) 3D basin stability profile of nodes from six-nodes motifs. Each axis represents a node's basin stability values at $K$=7, 14, and 21 on $x$, $y$, and $z$ axes, respectively. 
Nodes are clustered into four groups according to the node's basin stability at $K$ [$B(K)$]. The cluster $A$ ($C_A$) is illustrated as the purple cuboid for the nodes that are at the corner of the 3D space, whose  basin stabilities are $0 \leq B(7) \leq 0.2$, $0 \leq B(14) \leq 0.2$, and $0 \leq B(21) \leq 1$; the $C_B$ is the yellow cuboid for nodes with $0 \leq B(7) \leq 0.4$, $0.2 < B(14) \leq 1$, and $0.8 < B(21) \leq 1$; the green cuboid $C_C$ is the nodes with $0.4 < B(7) \leq 1$, $0.4 < B(14) \leq 1$, and $0.8 < B(21) \leq 1$; the $C_D$ shown as a blue cuboid is located in the middle of the 3D space that consists of nodes with $0.4 < B(7) \leq 0.8$, $0.4 < B(14) \leq 0.8$, and $0.4 < B(21) \leq 0.8$. 
The basic node or network characteristics such as degree (b) and edge density (c) do not show clear patterns, whereas betweenness (d) and flow betweenness (e) are roughly distributed from $C_A$ through $C_B$ and $C_C$ to $C_D$ as the values increase.} 
\label{fig_6}
\end{figure*}

In the 3D parameter space, the points are not randomly distributed (Fig.~\ref{fig_6}). The points (each point corresponds to each node) are clustered into four groups (cluster $C_A$, $C_B$, $C_C$, and $C_D$) as illustrated in Fig.~\ref{fig_6}(a). The nodes are clustered into groups according to the node's basin stability at $K$, which is denoted by $B(K)$. The cluster A ($C_A$) is illustrated as the purple cuboid for the nodes at the corner of the 3D space, whose basin stability values are $0 \leq B(7) \leq 0.2$, $0 \leq B(14) \leq 0.2$, and $0 \leq B(21) \leq 1$; the $C_B$ is the yellow cuboid for nodes with $0 \leq B(7) \leq 0.4$, $0.2 < B(14) \leq 1$, and $0.8 < B(21) \leq 1$; the green cuboid $C_C$ is the nodes with $0.4 < B(7) \leq 1$, $0.4 < B(14) \leq 1$, and $0.8 < B(21) \leq 1$; the $C_D$ shown as a blue cuboid is located in the middle of the 3D space that consists of nodes with $0.4 < B(7) \leq 0.7$, $0.4 < B(14) \leq 1$, and $0.4 < B(21) \leq 0.8$. The four clusters represent a handful of transition patterns. 

We compare the distribution of points with node characteristics such as degree, edge density of the corresponding network ensemble, betweenness, and flow betweenness as shown in Figs.~\ref{fig_6}(b)--~\ref{fig_6}(e). The position of nodes in the point cloud does not seem to have a clear correlation with degree and edge density. Some degree and edge density values are found at certain regions such as $K_7=0.5, K_{14}= 0.5, K_{21}=1$ for degree $=5$ or edge density $=1$ in Figs.~\ref{fig_6}(b) and ~\ref{fig_6}(c). However, it is hard to see any clear correlation for the entire point cloud. 

On the other hand, betweenness shows an interesting distribution on the points as shown in Fig.~\ref{fig_6}(d). We notice that the nodes in $C_A$ have higher betweenness (mean $=0.670$) than other nodes. The nodes can be easily distinguishable from the other nodes particularly at low $K$ values by their high betweenness. The nodes with the intermediate level of betweenness are clustered at $C_B$. The mean betweenness of the nodes in $C_B$ is $0.437$ and the basin stability is very high at $K=21$ for most of the nodes in cluster $C_B$ ($B>0.9$ for about 95\% of nodes in $C_B$ at $K=21$). The nodes in $C_B$ have small basin stability at $K=7$ similar to the high-betweenness nodes in $C_A$. However, the basin stability at $K=14$ is larger than that of the high-betweenness nodes in $C_A$ and widely distributed. 

It is interesting to note that there are two dominant clusters $C_C$ and $C_D$ where the low-betweenness nodes are located. One is a planar plane with basin stability $= 1$ at $K=21$ ($C_C$). Another dominant point is in the middle of the 3D space ($C_D$). The nodes in $C_D$ have the lowest betweenness (mean $= 0.033$), whereas the nodes belonging to $C_C$ have $0.085$ on average. The transition pattern seems to develop from the $C_A$ through $C_B$ and $C_C$ to $C_D$ area as the betweenness increases but it requires further investigation.

We compare the point cloud with our version of flow betweenness described in Sec.~\ref{sec:flow}. Since the flow betweenness specifically reflects the location of producer and consumer nodes, it provides clearer identification for the point cloud pattern than betweenness. For example, nodes in $C_A$ are distinguished from the other nodes with large flow betweenness values (mean $= 2.900$). Cluster $C_C$ is also well distinguished from $C_D$. The power of classification of betweenness and flow betweenness is visually noticeable in Fig.~\ref{fig_6}. The mean flow betweenness values of $C_B$, $C_C$, and $C_D$ are $1.566$, $1.088$, and $1.005$, respectively.

\subsubsection{\label{sec:k4} Abrupt stability increase and network characteristics}

One of the characteristics of the transition curves is the rate of basin stability increase as a function of $K$. Some nodes reach an intermediate level of basin stability at small $K \simeq 5$, but other nodes' basin stability values increase gradually as much as they have basin stability values $< 0.1$ at $K \simeq 20$ (see Fig.~\ref{fig_3}). The characteristics of the different rates for basin stability increase can be quantified by the $K$ value where the basin stability exceeds $0.4$ for the first time---$B(0.4)$---as we increase $K$ values.

\begin{figure}
\includegraphics[width=0.9\linewidth]{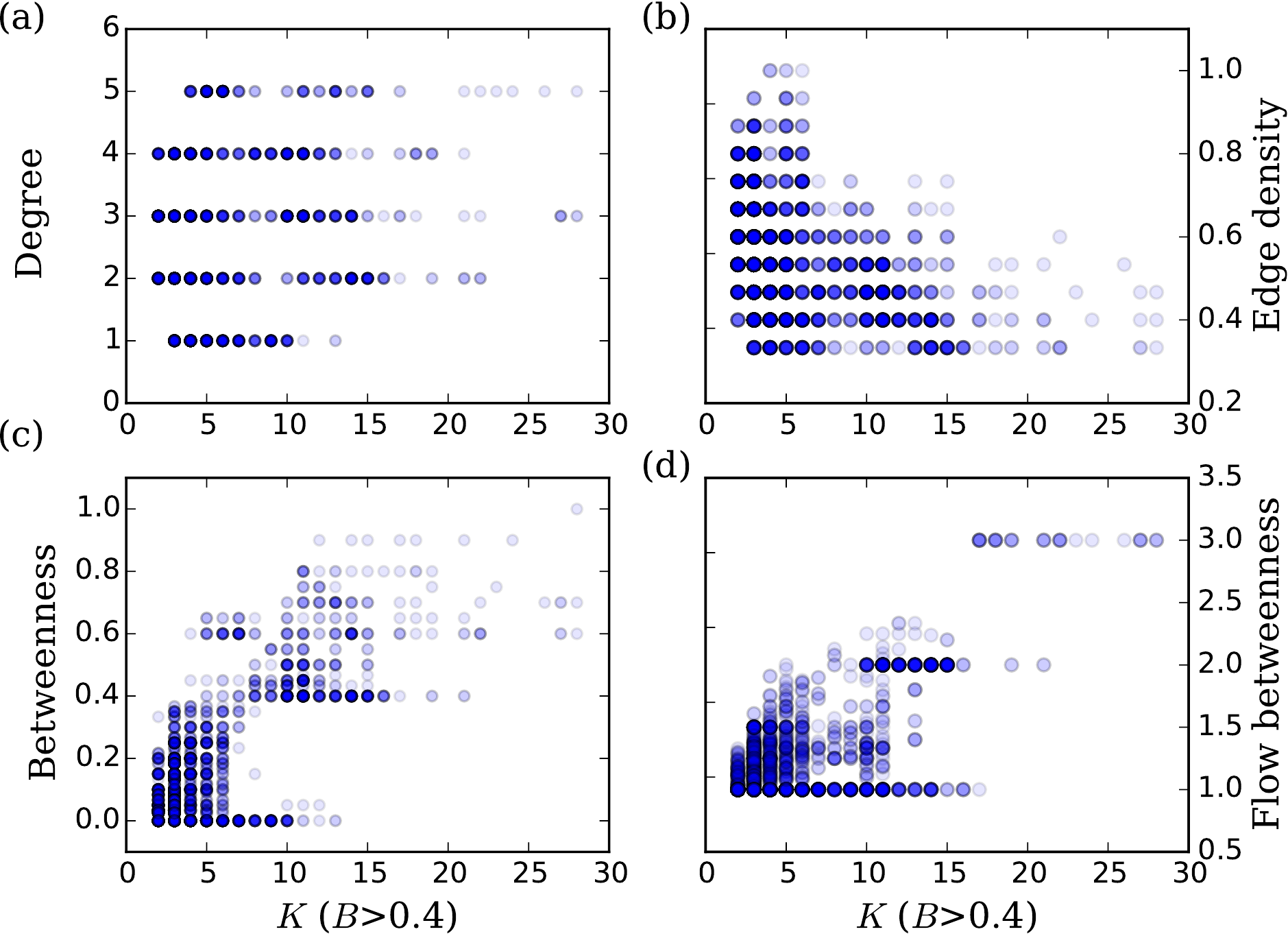}
\caption{(Color online) The correlation between the minimum coupling strength, where $B > 0.4$, and (a) degree, (b) edge density, (c) betweenness, and (d) current flow betweenness centralities.}
\label{fig_7}
\end{figure}

We compare $B(0.4)$ with the network characteristics such as degree, edge density, betweenness, and flow betweenness for the four-node network cases [Figs.~\ref{fig_7}(a), ~\ref{fig_7}(b), ~\ref{fig_7}(c), and ~\ref{fig_7}(d), respectively]. As we observe from the 3D point cloud, degree seems to have little to do with the transition shape. Edge density also does not show any clear correlation. However, it is interesting to see that the maximum edge density at each $B(0.4)$ seems to be negatively correlated with $B(0.4)$.

Betweenness and flow betweenness show stronger correlations between $B(0.4)$ than degree and edge density. The patterns of the scatter plots are very similar as the Pearson correlation coefficients are $0.749$ for betweenness and $0.732$ for flow betweenness. According to the correlation, the nodes with large betweenness or flow betweenness tend to have a low basin stability, which is in good agreement with Ref.~\cite{Schultz2014detours}. 

\section{\label{sec:discussions}Discussions and conclusion} 
In this study, we have investigated all of the unique network configurations for two-, four-, and six-node networks with two different node attributes of producer and consumer. We have revealed the basin stability transitions are classified into four clusters. Based on our analysis, we have shown that the transition patterns are not directly affected by either micro- or macro scale network characteristics. 
For example, the individual node degree is not correlated with the
transition patterns and the number of edges in a network is weakly
correlated with the transition patterns.
Instead, the transition pattern is affected by the collective interactions between nodes in a given network topology and node attributes. 
 
As an explanatory variable, we apply the betweenness in order to quantify the functional contribution of each node in a network. We also specify nodes' attributes and calculate flow betweenness with given producer-consumer pairs. Both betweenness and flow betweenness values are correlated with the transition patterns. Although flow betweenness is optimized for a power-grid system, betweenness shows the larger Pearson correlation value with the transition pattern than that of flow betweenness. 
Considering the statistical correlation to the transition patterns, flow betweenness does not improve betweenness in terms of summarizing the pattern of basin stability transition.

Our findings, in addition to our previous work~\cite{BasinStability_NJP}, have shown that the basin stability transition of nodes in networks is the result of a nontrivial combination of network topology and the position of the node in the network. We have provided some guidelines for simple but exhaustive possibilities of network configuration, which we hope helps one to investigate larger or more complex network structures with those simple structures as building blocks. 

There are other aspects of basin stability transition patterns worth investigating. For example, the temporary increase of basin stability appears, especially in dense network motifs. The detailed underlying mechanism can be addressed in the following research. 

\begin{acknowledgments}
This study is supported by Basic Science Research Program through the National Research Foundation of Korea (NRF) funded by the Ministry of Education (Grant No. 2013R1A1A2011947).
\end{acknowledgments}

\bibliography{reference}

\end{document}